\documentclass[aps,twocolumn,prd,superscriptaddress,
nofootinbib]{revtex4-2}

\usepackage[utf8]{inputenc}

\usepackage{mathtools}
\usepackage{amsfonts}
\usepackage{mathrsfs}
\usepackage{bbm}
\usepackage{physics}
\usepackage{slashed}
\usepackage{tensor}
\usepackage{setspace}
\usepackage{dsfont} % For the identity operator

\usepackage{graphicx}
\usepackage{color, float}
\usepackage{array}
\usepackage[abs]{overpic}

\usepackage{placeins}
\usepackage{booktabs}   % For clean rules
\usepackage{multirow}   % For \multirow command
\usepackage{makecell}
\usepackage{subcaption}
\usepackage{xspace}
\usepackage{siunitx}
\usepackage{xfrac}
\usepackage{hyperref}
\usepackage[nameinlink]{cleveref}
\usepackage{appendix}
\def \W{\mathcal{W}}

\newcommand{\ttt}[1]{\texttt{#1}} % Comandos do Mathematica

\usepackage{xifthen}
\usepackage{xcolor}
\hypersetup{
	colorlinks,
	linkcolor={red!75!black},
	citecolor={blue!75!black},
	urlcolor={blue!75!black}
}

\usepackage{booktabs}
\usepackage{multirow}

\newcolumntype{C}{>{$}c<{$}}
\AtBeginDocument{
	\heavyrulewidth=.08em
	\lightrulewidth=.05em
	\cmidrulewidth=.03em
	\belowrulesep=.65ex
	\belowbottomsep=0pt
	\aboverulesep=.4ex
	\abovetopsep=0pt
	\cmidrulesep=\doublerulesep
	\cmidrulekern=.5em
	\defaultaddspace=.5em
}

\graphicspath{{./figures/}}

\newcommand{\gettitle}{Mass Dependence of the Araki-Uhlmann Relative Entropy Across Dimensions
	}

\newcommand{\getUERJAffiliation}{\affiliation{UERJ $–$ Universidade do Estado do Rio de Janeiro,	Instituto de Física $–$ Departamento de Física Teórica $–$ Rua São Francisco Xavier 524, 20550-013, Maracanã, Rio de Janeiro, Brazil}}
\newcommand{\getCBPFffiliation}{\affiliation{CBPF $-$ Centro Brasileiro de Pesquisas Físicas, Rua Dr. Xavier Sigaud 150, 22290-180, Rio de Janeiro, Brazil}}
\newcommand{\getUFRJAffiliation}{\affiliation{UFRJ $-$ Universidade Federal do Rio de Janeiro, Instituto de Física, RJ 21.941-972, Brazil}}
\newcommand{\getUFRRJffiliation}{\affiliation{UFRRJ $-$ Universidade Federal Rural do Rio de Janeiro, Departamento de Física, Zona Rural - BR-465, Km 07, 23890-000, Seropédica, Rio de Janeiro, Brazil}}

\hypersetup{
	pdftitle={\gettitle},
	bookmarksopen=true,
	bookmarksopenlevel=2,
	bookmarksnumbered=true
}

\begin{document}
	
	\title{\gettitle}
	
	\author{João G. A. Caribé}
    \email{joaogcaribe@ufrrj.br}
	\getUFRRJffiliation
	\author{Marcelo S. Guimarães}
    \email{msguimaraes@uerj.br}
	\getUERJAffiliation
	\author{Itzhak Roditi}
    \email{roditi@cbpf.br}
	\getCBPFffiliation
	\author{Silvio P. Sorella}
    \email{silvio.sorella@fis.uerj.br}
	\getUERJAffiliation
	\author{Arthur F. Vieira}
    \email{arthurfvieira@if.ufrj.br}
	\getUERJAffiliation\getUFRJAffiliation
	\begin{abstract}
	We investigate the mass dependence of the Araki-Uhlmann relative entropy between a localized coherent excitation and the vacuum state of a free scalar quantum field on the $(1+d)$-dimensional Minkowski spacetime for $d=1,2,3$. In this context, the relative entropy admits a closed expression in terms of the smeared Pauli-Jordan distribution, whose analytic structure is sensitive to both the mass and the spacetime dimensionality. Prior studies in $(1+1)$ dimensions have shown a monotonic decay of the relative entropy with increasing mass. We extend  that analysis to higher dimensions using numerical techniques and elucidate how the interplay between dimensionality and mass controls the behavior of the relative entropy. Our results provide new insights for the study of the Araki-Uhlmann relative entropy in QFT and its dependence on physical parameters. 		
	\end{abstract}
	\maketitle

	%%%%%%%%%%%%%%%%%%%%%%%%%%%%%%%%%
	\section{Introduction} 
	\label{sec:Introduction} 
	The study of entanglement in quantum field theory (QFT) became central to diverse areas ranging from quantum information and condensed matter physics to black hole thermodynamics and holography \cite{Nishioka:2018khk,Holzhey:1994we,Calabrese:2004eu,Calabrese:2009qy,Casini:2022rlv,Berges:2017hne,Schrofl:2023hnz,Katsinis:2023hqn}. A primary tool for quantifying correlations in quantum systems is the von Neumann entropy, which upon reduction across a partition of a bipartite quantum system gives rise to the notion of entanglement entropy. In relativistic QFT, the entanglement entropy between spatial regions captures fundamental features of quantum correlations but is plagued by ultraviolet divergences due to short-distance degrees of freedom near the entangling surface \cite{Casini:2022rlv,Marolf:2016dob}. The resulting entropy typically exhibits a divergent structure consisting of power-law and logarithmic terms, with only the logarithmic coefficient being universal. These divergences, largely state-independent, underscore the need for better-behaved quantities to probe the structure of entanglement.
	
	A natural refinement is provided by the relative entropy between two quantum states \cite{Hiai:1991mxv,Casini:2022rlv}. In a finite-dimensional system, relative entropy measures the distinguishability between two of its states, is non-negative and vanishes if and only if the two states coincide. Importantly, in QFT the ultraviolet divergences cancel in the relative entropy, rendering it finite and physically meaningful. Relative entropy has since found applications in thermodynamics \cite{Floerchinger:2020ogh,Dowling:2020nxc}, algebraic QFT \cite{Hollands:2019czd,DAngelo:2021yat,Ciolli:2021otw,Galanda:2023vjk,Garbarz:2022wxn}, and gauge-gravity duality \cite{Jafferis:2014lza,Jafferis:2015del,Verlinde:2019ade,Bousso:2020yxi}.

    The algebraic framework of Tomita-Takesaki modular theory \cite{tomita1967canonical,Takesaki:1970aki} elevates these notions to a fully covariant setting. Within this context, Araki and Uhlmann \cite{Araki:1975zw,Araki:1976zv,Uhlmann:1976me} introduced a generalization of relative entropy for von Neumann algebras, now known as the Araki-Uhlmann relative entropy. For a matter of convenience, we will refer to it simply as relative entropy throughout this text. It is defined for two states $\ket{\Psi}$ and $\ket{\Omega}$ as
	\begin{equation}\label{ArakiS}
		S(\Psi|\Omega)=-\bra{\Psi}\log \Delta_{\Psi\Omega}\ket{\Psi},
	\end{equation}
    where $\Delta_{\Psi\Omega}$ is the relative modular operator, and satisfies a number of important structural properties. Most notably, it is positive definite \cite{Witten:2018zxz}:
    \begin{equation}
    	S(\Psi|\Omega)\geq 0,
    \end{equation}
    with equality if and only if $\ket{\Psi}=\ket{\Omega}$. This reflects the basic physical intuition that distinct states must differ in their informational content. Additionally, it obeys monotonicity under restriction to subalgebras, often referred to as the data processing inequality. If $\mathcal{O}_1 \subset \mathcal{O}_2$ are nested spacetime regions, then
    \begin{equation}
    	S_{\mathcal{O}_1}(\Psi|\Omega)\leq S_{\mathcal{O}_2}(\Psi|\Omega),
    \end{equation}
    indicating that the distinguishability between two states cannot increase upon reducing the set of accessible observables. The relative entropy also enjoys further well-established features, such as joint convexity, additivity over tensor product states, and invariance under unitary evolution \cite{Witten:2018zxz}.

    In $(1+1)$ dimensions, previous studies enabled by the availability of analytic modular Hamiltonians have shown that the relative entropy decays with increasing mass \cite{Guimaraes:2025cqt} (see also \cite{Caribe:2025knm}). This is consistent with the physical intuition that heavier fields exhibit shorter-range correlations, reducing the information-theoretic distinguishability between coherent excitations and the vacuum. However, analogous analyses in higher dimensions remain scarce due to the increased complexity of the modular structure and the lack of closed-form expressions.
    
    This work aims to fill this gap by studying the relative entropy in $(1+2)$ and $(1+3)$ dimensions via numerical evaluation of Eq.~\labelcref{ArakiSPJ}. The methodology relies on using the Quasi-Monte Carlo method to compute the integrals involving the Pauli-Jordan distribution, together with Lorentz covariance and localization properties of the coherent excitation. Our goal is to quantify how the relative entropy responds to variations in the mass of the field and the dimensionality of the spacetime, providing new insight into the structure of modular theory in higher-dimensional QFT.

    The key insight of this work lies in the fact that the relative entropy between a coherent state and the vacuum state of a massive free scalar field on the Minkowski spacetime depends on the mass and spacetime dimensionality entirely through the Pauli-Jordan distribution. This simplification facilitates the systematic numerical investigation of how the scalar field mass and the spacetime dimensionality affect the distinguishability between these states.
    
    The rest of this paper is organized as follows: \Cref{sec:Setup} discusses the physical setup of this work and is further divided into three subsections: ~\Cref{sec:PJinDdims},~\Cref{sec:test_Functions} and~\Cref{sec:Numerical_scheme} that present the Pauli-Jordan distribution in different dimensions, the test function we employ and the numerical strategy for evaluating Eq.~\labelcref{ArakiSPJ}, respectively.  \Cref{sec:Results} presents our results on the mass and spacetime dimensionality dependence of the Araki-Uhlmann relative entropy. \Cref{sec:Conclusions} offers conclusions and directions for future work.

    %%%%%%%%%%%%%%%%%%%%%%%%%%%%%%%%%%%%%%%
    \section{Setup}\label{sec:Setup}
    On the $(1+d)$-dimensional Minkowski spacetime we consider the case where $\ket{\Omega}$ is the vacuum of a free massive scalar field and $\ket{\Psi}=\ket{\psi_f}$ is a coherent excitation localized in the right Rindler wedge $W_{\rm R}$. This region is given by 
    \begin{equation}\label{eq:Right_Rindler_Wedge}
	 	W_{\rm R}=\left\{(t,x^1,\hdots,x^d) \in \mathbb{R}^{1+d}\,|\,\, x^1 > |t|\right\}
	  \end{equation}
    in the Minkowski coordinates~$\{t\in\mathbb{R},x^{1}\in\mathbb{R},\hdots,x^{d}\in\mathbb{R}\}$ of a suitable reference frame. The coherent state $\ket{\psi_f}$ is given by the action of the Weyl operator $e^{i\varphi(f)}$ on the vacuum state:
	\begin{equation}\label{eq:defintion of coherent state}
		\ket{\psi_f}=e^{i\varphi(f)}\ket{\Omega},
	\end{equation}
	where
	\begin{equation}\label{eq: Smeared field}
		\varphi(f)=\int \dd^{1+d}\vb*{x}\, f(\vb*{x})\,\varphi(\vb*{x})
		\end{equation}
    is the smeared field operator, $f\in\mathcal{C}_0^{\infty}(W_{\rm R})$ is a real-valued smooth (infinitely differentiable) test function that is compactly supported in $W_{\rm R}$, $\vb*{x} = (t,\vec{x})$, and
    \begin{equation}
        \varphi(\vb*{x}) = \frac{1}{(2\pi)^d}\int\dd^{d}\vec{k}\frac{1}{2\omega_k}\left(e^{-ik_\mu x^{\mu}}a_{k} + e^{ik_\mu x^{\mu}}a^{\dagger}_{k}\right)
    \end{equation}
    is the field operator-valued distribution, where  $k_\mu x^{\mu} = \omega_k t - \vec{k}\cdot\vec{x}$, $\omega_k = \sqrt{\vec{k}^2 + m^2}$ and $a_k$ and $a^{\dagger}_k$ are, respectively, the annihilation and creation operators that obey the canonical commutation relations
    \begin{equation}
        [a_k, a^{\dagger}_{k'}] = (2\pi)^d \, 2\omega_{k} \, \delta^{d}(\vec{k} - \vec{k}')\mathds{1}
    \end{equation}
    and
    \begin{equation}
        [a_k, a_k] = [a^{\dagger}_k, a^{\dagger}_{k'}] = 0.
    \end{equation}

    In this setup, the relative entropy between $\ket{\psi_f}$ and $\ket{\Omega}$ is considerably simplified. In particular, it can be written 
    % in terms of a Lorentz-boosted test function in the direction of $x^1$ and the Pauli–Jordan commutator distribution 
    as \cite{Casini:2019qst,Frob:2024ijk}
   	\begin{equation}\label{ArakiSPJ}
		S(\psi_f|\Omega)=-\frac{1}{2}\Delta_{\rm PJ}(f,f'_s|_{s=0}),
	\end{equation}
	where $\Delta_{\rm PJ}$ is the smeared Pauli-Jordan distribution, $f_s(\vb*x)=f(\Lambda_{-s}\vb*x)$ is the Lorentz-boosted profile of the test function $f(\vb*{x})$ under the one-parameter boost family $\Lambda_s$ given by
	\begin{align}
		\Lambda_s x^{\mu}: 
		\left\{
		\begin {aligned}
		&  \;\;  t'  =  \cosh(2\pi s) \;t - \sinh(2 \pi s) \; x^1,\\
        &  \;\; x'^{1}  =  \cosh(2\pi s) \;x^1 - \sinh(2 \pi s) \; t, \\
		&\;\; x'^{2}=x^2,\\
        &\;\;\;\;\;\;\;\;\, \vdots\\
		&\;\; x'^{d}=x^{d},\\          
	\end{aligned}
	\right. \label{bst}
	\end{align}
	and $f'_s(\vb*x)=\frac{\partial}{\partial s}f(\Lambda_{-s}\vb*x)$.
    
The smeared Pauli-Jordan distribution is defined as
\begin{equation}\label{eq:Smeared_PJ}
\Delta_{\rm PJ}(f,g)=\int\dd^{1+d}\vb*{x}\,\dd^{1+d}\vb*{y}f(\vb*{x})\Delta_{\rm PJ}^{(1+d)}(\vb*{x}-\vb*{y})g(\vb*{y}),
\end{equation}
where the Pauli-Jordan distribution in $(1+d)$ dimensions is given by the antisymmetric part of the two-point Wightman function \cite{Bogolyubov:1959bfo,Tjoa:2021roz},
	\begin{equation}\label{PJDef}
	\Delta_{\rm PJ}^{(1+d)}(\vb*{x}-\vb*{x}')=\frac{\W^{(1+d)}(\vb*{x}-\vb*{x}')-\W^{(1+d)}(\vb*{x}'-\vb*{x})}{2i},
\end{equation}
with
 \begin{equation}\label{Wightman}
	\W^{(1+d)}(\vb*{x}-\vb*{x}')=\int \frac{\dd^d\vec{k}}{(2\pi)^d}\frac{1}{2 \omega_k}e^{-ik_\mu(x^{\mu}-x'^{\mu})},
\end{equation}
where $m$ is the mass and $\vb*{x'}=(t',x'^1,\hdots,x'^d)$.

Expressions \labelcref{ArakiS} and \labelcref{ArakiSPJ} are not merely formal constructs, but also encode deep structural features of QFT, including locality and causality. In particular, they are intimately connected to the algebraic formulation of QFT via the Haag–Kastler axioms. In wedge regions, the Bisognano-Wichmann theorem links the modular flow of the vacuum to Lorentz boosts, providing a concrete realization of the modular operator in relativistic settings, see \cite{Witten:2018zxz} for a detailed account. Moreover, these expressions reflect the Reeh-Schlieder property, which ensures the cyclic and separating nature of the vacuum with respect to local algebras, and are related to the structure of entanglement wedges holography \cite{Nishioka:2018khk}.
	
	%%%%%%%%%%%%%%%%%%%%%%%%%%%%%%%%%%
	\subsection{The Pauli-Jordan distribution in $(1+d)$ dimensions with $d=1, 2, 3$} 
	\label{sec:PJinDdims}
	Given that Eq.~\labelcref{PJDef} defines the Pauli-Jordan function as the antisymmetric part of the Wightman two-point function in $(1+d)$ dimensions, explicit closed-form expressions can be obtained in the cases $(1+1)$, $(1+2)$, and $(1+3)$, yielding the following expressions \cite{Bogolyubov:1959bfo,Tjoa:2021roz,gradshteyn2014table}: 
    \begin{subequations}
	\begin{align}
			\Delta_{\rm PJ}^{(1+1)}(\vb*{x})&=-\frac{1}{2}\textmd{sign}(t)\Theta(t^2-x^2)J_0(m\sqrt{t^2-x^2}),\label{eq::Paulis1}\\
			\Delta_{\rm PJ}^{(1+2)}(\vb*{x})&=\frac{1}{4 \pi}\textmd{sign}(t)\frac{\Theta(t^2-\vec{x}^{\,2})}{\sqrt{t^2-\vec{x}^{\,2}}}\sin(m\sqrt{t^2-\vec{x}^{\,2}})\label{eq::Paulis2}\\
			\Delta_{\rm PJ}^{(1+3)}(\vb*{x})&=-\frac{1}{4\pi}\textmd{sign}(t)\delta(t^2-\vec{x}^{\,2})\nonumber\\
            &+\frac{m}{8\pi}\textmd{sign}(t)\frac{\Theta(t^2-\vec{x}^{\,2})}{\sqrt{t^2-\vec{x}^{\, 2}}}J_1(m\sqrt{t^2-\vec{x}^{\,2}}),\label{eq::Paulis3}
		\end{align}
        \end{subequations}
	where $J_0(\xi)$ and $J_1(\xi)$ are the zeroth- and first-order Bessel function of the first kind respectively, and $\Theta(\vb*{x})$ is the Heaviside step function. For large arguments, $\xi \gg 1$, the Bessel functions admit the asymptotic expansions~\cite{NIST:DLMF}	\begin{align}
		J_0(\xi)&\sim \sqrt{\frac{2}{\pi \xi}}\cos{\left(\xi-\frac{\pi}{4}\right)+\mathcal{O}(\xi^{-3/2}) },\\
		J_1(\xi)&\sim \sqrt{\frac{2}{\pi \xi}}\cos{\left(\xi-\frac{3\pi}{4}\right)+\mathcal{O}(\xi^{-3/2}) }.
	\end{align}
    
	From these expressions it follows that the large-$m$ behavior of the Pauli-Jordan functions in Eqs.~\labelcref{eq::Paulis1}, \labelcref{eq::Paulis2}, and \labelcref{eq::Paulis3} for timelike separated points are characterized by an oscillatory behavior, with dimensionality dictating distinct qualitative patterns: in $(1+1)$ dimensions the amplitude of the oscillations decays as $m^{-1/2}$; in $(1+2)$ dimensions, it exhibits rapid oscillations with non-decaying (with increasing mass) amplitude; and in $(1+3)$ dimensions, the amplitude of the oscillations increase as $m^{1/2}$. Consequently, the dependence of the Araki-Uhlmann relative entropy on the mass parameter is strongly dimension-sensitive, reflecting how spacetime dimensionality shapes the structure of quantum correlations in relativistic field theory.

	%%%%%%%%%%%%%%%%%%%%%%%%%%%%%%%%%%
	\subsection{Test functions} 
	\label{sec:test_Functions}
    In this work we consider a separable test function $f\in\mathcal{C}_0^{\infty}(W_{\rm R})$, i.e., that can be written as
    \begin{equation}
		f(\vb*{x})=\eta f_t(t)f_x(x)f_y(y)f_z(z).
	\end{equation}
    Except for the normalization constant $\eta\in\mathbb{R}$, each of its factors is also assumed to be an element of $\mathcal{C}_0^{\infty}(W_{\rm R})$. One has to choose a test-function in order to compute the relative entropy with Eq.~\labelcref{ArakiSPJ}. However, such choice should not affect qualitative behavior of the mass-dependence of the relative entropy. Evidence for this conjecture in a similar setup is provided by~\cite{Guimaraes:2025cqt} which showed that the qualitative behavior of the relative entropy with respect to the field's mass is the same for different choices of test functions. Given this conjecture, our choice of a separable test function is justified by a matter of simplicity: Its behavior is easier to understand as it does not couple the different coordinates of $\vb*{x}$. 
    
    As the building block of that separable test function we use the bump function~\cite{Nestruev_2020}
    \begin{equation}
        \mathrm{B}(x)=
            \begin{cases}
                e^{\left(\dfrac{1}{x^2-1}\right)}, \quad \abs{x} < 1\\
                0, \quad \text{otherwise},
            \end{cases}
    \end{equation}
    which is infinitely differentiable and compactly supported in the region $\abs{x}< 1$. With that, we define our smooth and compactly supported test function as 
    \begin{equation}\label{eq:test_function}
        f(\vb*{x})=
        \eta \mathrm{B}(t/\alpha)\mathrm{B}((x-l)/\beta)\mathrm{B}(y/\beta)\mathrm{B}(z/\beta)
    \end{equation}
    for the $(1+3)$-dimensional case. Its support lies on a cuboid-shaped region with timelike edges of length $2\alpha > 0$ and spacelike edges of length $2\beta > 0$. Both of these are defined with respect to a static observer. The parameter $l$ controls the spacelike distance between the center of the cuboid and the worldline of a static observer at $x = 0$. 

    The $(1+1)$- and $(1+2)$-dimensional versions of that test function are obtained by suitably removing its $y$- and/or $z$-dependent factors. In all of these cases, we ensure that $f$ is compactly supported in $W_{\rm R}$ by choosing $\alpha, \beta$ and $l$ such that 
    \begin{equation}
        l-\beta > \alpha.
    \end{equation}
    
    To compute the relative entropy it is also necessary to evaluate $f'_s|_{s=0}$ (Eq.~\labelcref{ArakiSPJ}). For our choice of test function this calculation is straightforward and the resulting expression is
    \begin{widetext}
        \begin{equation}\label{eq:test_function_boosted}
            f'_s|_{s=0}(\vb*{x}) = -4 \pi t \left(\frac{(x-l)}{\beta^2}\left(\frac{1}{((x-l)/\beta)^2-1}{}\right)^2+\frac{x}{\alpha^2}\left(\frac{1}{(t/\alpha)^2-1}\right)^2\right)f(\vb*{x}).
        \end{equation}
    \end{widetext}

\begin{figure*}
		\centering%
		\includegraphics[width=\textwidth]{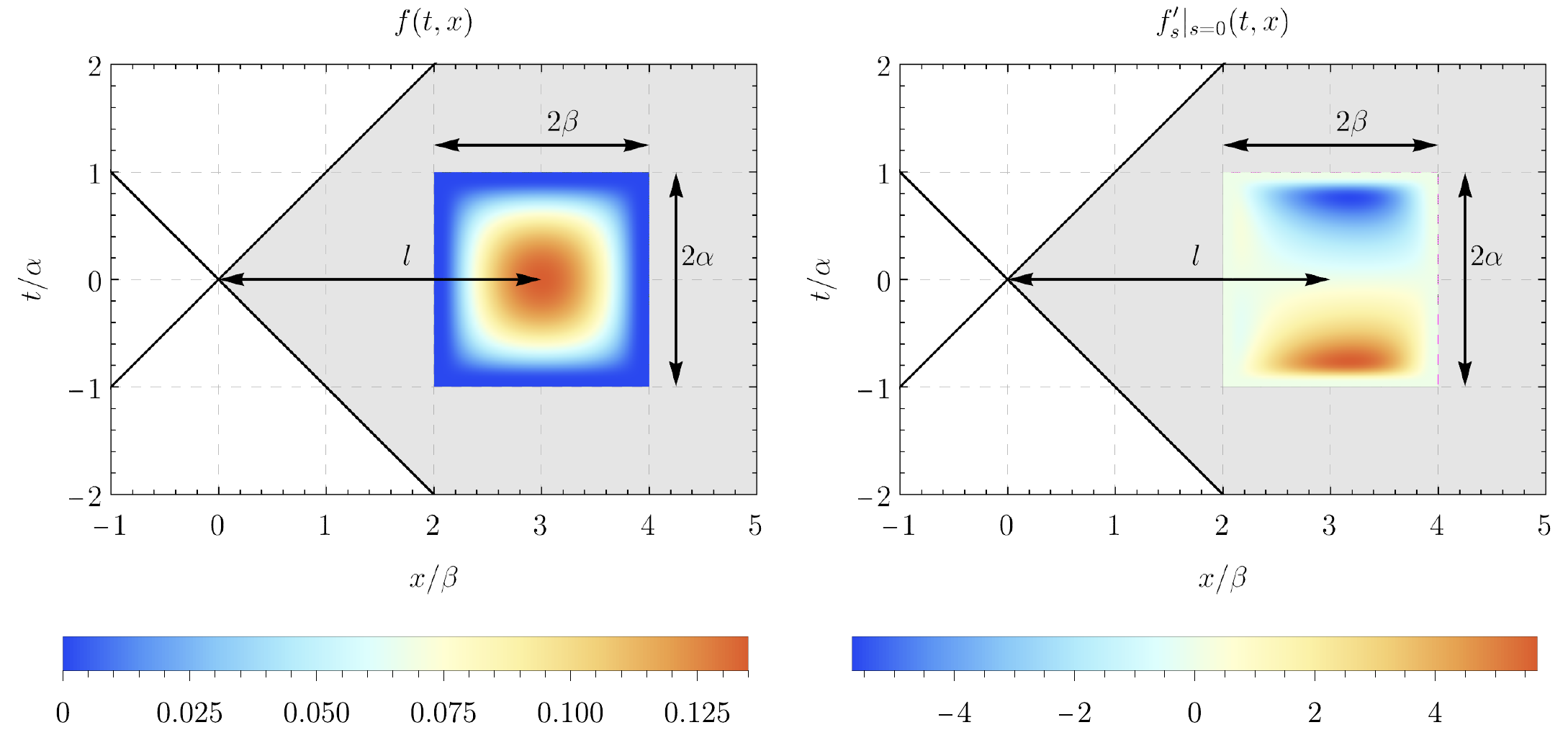}
		\caption{Density plot of $f$ and $f'_s|_{s = 0}$ from Eqs.\labelcref{eq:test_function} and~\labelcref{eq:test_function_boosted}, respectively, for the $(1+1)$-dimensional case. In this plot, $\alpha = 1 = \beta$ and $l = 3$. As can be seen, both are compactly supported in the right Rindler wedge $W_{\rm R}$ defined in Eq.\labelcref{eq:Right_Rindler_Wedge} and represented here as the gray shaded region.}
		\label{fig:Test_function}
	\end{figure*}

    For the convenience of the reader, the $(1+1)$-dimensional versions of $f$ and $f'_s|_{s=0}$ are presented in ~\Cref{fig:Test_function} for $\alpha = \beta = 1$ with $l = 3$ which are the values we use throughout this work.
        
    To proceed with the evaluation of the relative entropy from Eq.\labelcref{ArakiSPJ}, one has to solve the $2(1+d)$-dimensional spacetime integral that results from plugging $f$ and $f'_s|_{s=0}$ in Eq.~\labelcref{eq:Smeared_PJ} for the smeared Pauli-Jordan distribution. This integral depends on the mass $m$ of the field through the Pauli-Jordan distribution. The direct analytical evaluation of this expression is extremely difficult. While the Pauli-Jordan distribution 
	$\Delta^{(1+d)}_{\rm PJ}(\vb*{x}-\vb*{x}';m)$ has a simple form in momentum space, the Fourier transform of the test-functions $f$ and $f'_s|_{s=0}$ from equations~\labelcref{eq:test_function} and~\labelcref{eq:test_function_boosted}, respectively, do not have a closed-form analytical expression. Consequently, transforming the problem to the momentum space would not simplify the calculation. Instead, it would increase the dimensionality of the integral, resulting in an even greater computational challenge. Therefore, the computation is carried out directly in position space using the numerical procedure presented in what follows.

    %%%%%%%%%%%%%%%%%%%%%%%%%%%%%%%%%%%%%%%
    \subsection{Numerical procedure}\label{sec:Numerical_scheme}   
    To evaluate the $2(1+d)$-dimensional integral that appears in the smeared Pauli-Jordan distribution, we use the \ttt{QuasiMonteCarlo} method from Mathematica~\cite{Mathematica}. As described in Sec.~\labelcref{sec:test_Functions}, we consider $\eta = 1$, $\alpha = 1 = \beta$ and $l = 3$ to setup the test function from Eq.~\eqref{eq:test_function}. While the absolute value of the relative entropy depends on this choice, its qualitative dependence on the field mass $m$ is expected to be universal in the limits of small and large $m\alpha$, as these regimes probe the long- and short-distance behavior of the theory. The universality of that qualitative dependence in the intermediary mass regime is conjectured in~\cite{Guimaraes:2025cqt}, where several different test functions produced relative entropies with the same qualitative dependence on the field mass $m$.

    Within that setup, we probe the mass range $m\alpha\in[10^{-10},20]$, which spans all the relevant physical regimes. The low-mass regime occurs when the Compton wavelength $m^{-1}$ is much larger than the smearing scale $\sim 2\alpha$, and the large-mass regime happens in the opposite limit. As we have chosen $\alpha = \beta = 1$, the low- and large-mass regimes in our case are characterized by $2m\alpha \ll 1$ and $2m\alpha \gg 1$,  respectively. Therefore, the range of masses we considered is sufficient to probe all mass regimes. The mass parameter was sampled logarithmically for $m\alpha\in[10^{-10},10^{-1}]$ and linearly with step $\Delta m\alpha = 10^{-1}$ for $m\alpha \in\; (10^{-1},20]$. The massless case ($m = 0$) was also considered for $(1+2)$ and $(1+3)$ spacetime dimensions.

    The \texttt{QuasiMonteCarlo} method in this context is subject to two main sources of error: the one related to dimensionality, which increases the absolute error for a fixed number of samples as the dimensionality of the integral increases, and the challenge of resolving the decaying relative entropy in the high-mass regime, which requires a small absolute error to maintain a low relative error. Both of these issues are mitigated by increasing the number of samples over which the \texttt{QuasiMonteCarlo} integration is performed. However, we can not increase this number arbitrarily because the computational time it takes to run the integration scales linearly with the number of sampling points. A practical upper limit for sampling points in this computation was set to $10^8$, as this typically resulted in a computation time of less than $1$ day per batch on our hardware. Using $10^9$ points causes a single integration to take roughly one day, rendering the full parameter sweep computationally prohibitive. To manage this trade-off systematically, we performed a convergence analysis for representative mass values in each spacetime dimension. We determined the minimum number of sampling points (MaxPoints) required to achieve convergence of the integral to within a specified PrecisionGoal. The resulting integration parameters are summarized in \Cref{tab:integration_params}.
    
    \begin{table}[h!]
        \centering
        \begin{tabular}{l l c c}
            \toprule
            \textbf{Dimensions} & \multicolumn{1}{c}{$m\alpha$} & \textbf{\texttt{PrecisionGoal}} & \textbf{\texttt{MaxPoints}} \\
            \midrule
            \multirow{3}{*}{(1+1)} & \( [10^{-10}, 4) \) & 2 & \( 10^7 \) \\
                                   & \( [4, 10) \)       & 2 & \( 10^8 \) \\
                                   & \( [10, 20] \)      & 3 & \( 10^8 \) \\
            \midrule
            \multirow{2}{*}{(1+2)} & \( [10^{-10}, 3) \) & 3 & \( 10^7 \) \\
                                   & \( (3, 20] \)       & 3 & \( 10^8 \) \\
            \midrule
            (1+3) & \( [10^{-10}, 20] \) & 3 & \( 10^8 \) \\
            \bottomrule
        \end{tabular}
        \caption{Numerical integration parameters. Parameters were determined via convergence studies to guarantee the \texttt{PrecisionGoal}.}
        \label{tab:integration_params}
    \end{table}

    Finally, it is important to remark that in $(1+3)$ spacetime dimensions, the Pauli-Jordan distribution (Eq.~\labelcref{eq::Paulis3})
    contains a term supported on the light cone, $\propto \delta (t^2 - \vec{x}^2)$. A direct numerical integration of the full expression fails to capture this singular contribution. We therefore isolate it by decomposing the smeared Pauli-Jordan distribution as 
    \begin{equation}\label{eq:decomp-PJ}
                    \Delta_{\mathrm{PJ}}(f,g) = \Delta^{\mathrm{LC}}_{\mathrm{PJ}}(f,g) + \Delta^{\mathrm{BULK}}_{\mathrm{PJ}}(f,g),
    \end{equation}
    where $\Delta^{\mathrm{LC}}_{\mathrm{PJ}}(f,g)$ is the contribution that arises from the light cone and $\Delta^{\mathrm{BULK}}_{\mathrm{PJ}}(f,g)$ is the contribution that arises from the bulk of the light cone. This decomposition is essential. We found that the bulk term alone yields a negative contribution, which would lead to a violation of the positive-definiteness of the relative entropy. The light-cone term provides the dominant positive contribution in the low-mass regime necessary to recover a physically consistent result.   
    
    We compute the light cone term by first integrating over the $\delta$-function analytically (see App.~\labelcref{App: The light cone contribution to the PJ}), which reduces the integral's dimensionality. The resulting integral is then computed using the \texttt{QuasiMonteCarlo} method with \ttt{PrecisionGoal} = 4 and \ttt{MaxPoints} = $10^9$ which results in a relative error of order $\sim 10^{-3}$ as reported by the Mathematica error estimator for the \texttt{QuasiMonteCarlo} method. The bulk term $\Delta^{\mathrm{BULK}}_{\mathrm{PJ}}(f,g)$, involving $\Theta(t^2 - \vec{x}^2)$ is computed using the protocol delineated in the previous paragraph.
        
	%%%%%%%%%%%%%%%%%%%%%%%%%%%%%%%%%%
	\section{Results} 
	\label{sec:Results}
    \begin{figure*}
		\centering%
		\includegraphics[width=\textwidth]{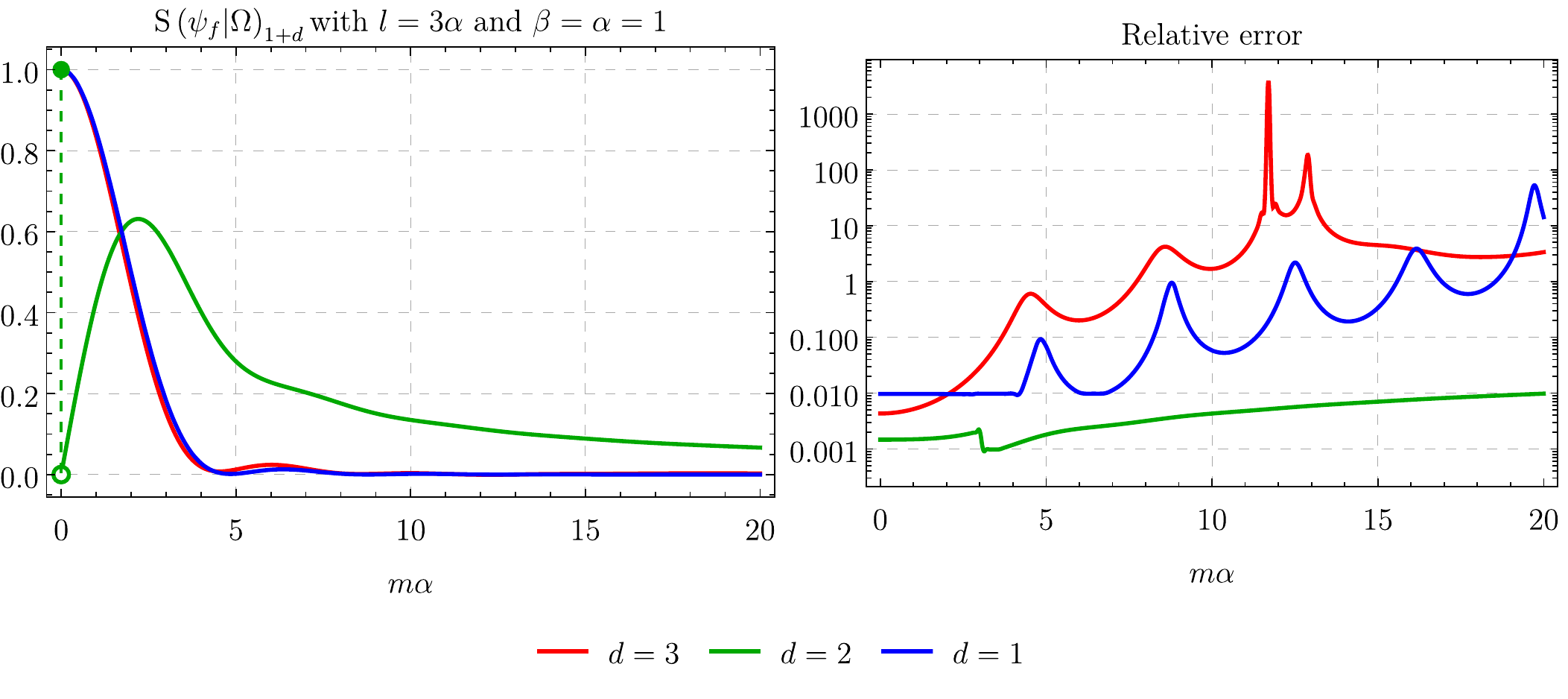}
		\caption{The relative entropy $S(\psi_f|\Omega)$ between $\ket{\psi_f}$ and $\ket{\Omega}$ as a function of the mass $m\alpha$ of the field on a $(1+d)$-dimensional Minkowski spacetime and the corresponding relative errors are presented on the left and right plots, respectively. Here $\alpha = 1 = \beta$ and $l = 3\alpha$. The red, green and blue curves represent the $d = 3$, $d = 2$ and $d = 1$ cases in both plots.
        In each case, $\eta$ is chosen such that the entropies peak at $1$. For $d = 1$ and $d = 3$ the entropies peak as $m \to 0$ (for $d = 1$, the massless case $m = 0$ is unphysical due to the infrared divergence of the theory in this case) and decays with increasing mass with dwindling oscillations in the process. For $d = 2$, the relative entropy is minimum as $m\to 0$ (up to the discontinuity at $m = 0$), increases towards a maximum near $m\alpha \approx 2.2$ then decays slower than in the $d=1$ and $d = 3$ cases. The $d=2$ case also presents an oscillation while decaying, which is dampened faster than those for the $d=1$ and $d=3$ cases. For this reason, we can only see a trace of an oscillation around $m=7$ in the green-curve on the left plot. The oscillations in the relative error follow the oscillations in the relative entropy. When the relative entropy becomes closer to zero, the absolute error becomes more pronounced, which produces the oscillations we see on the right plot. With the evaluation settings we used, the absolute error stabilizes around $10^{-5}$, $10^{-3}$ and $10^{-2}$ for $d = 1, 2$ and 3, respectively. Concomitantly, the relative entropy decays with increasing mass. For this reason, the relative error grows with increasing mass.}
		\label{fig:Results_1}
	\end{figure*}

     \begin{figure*}
		\centering%
		\includegraphics[width=\textwidth]{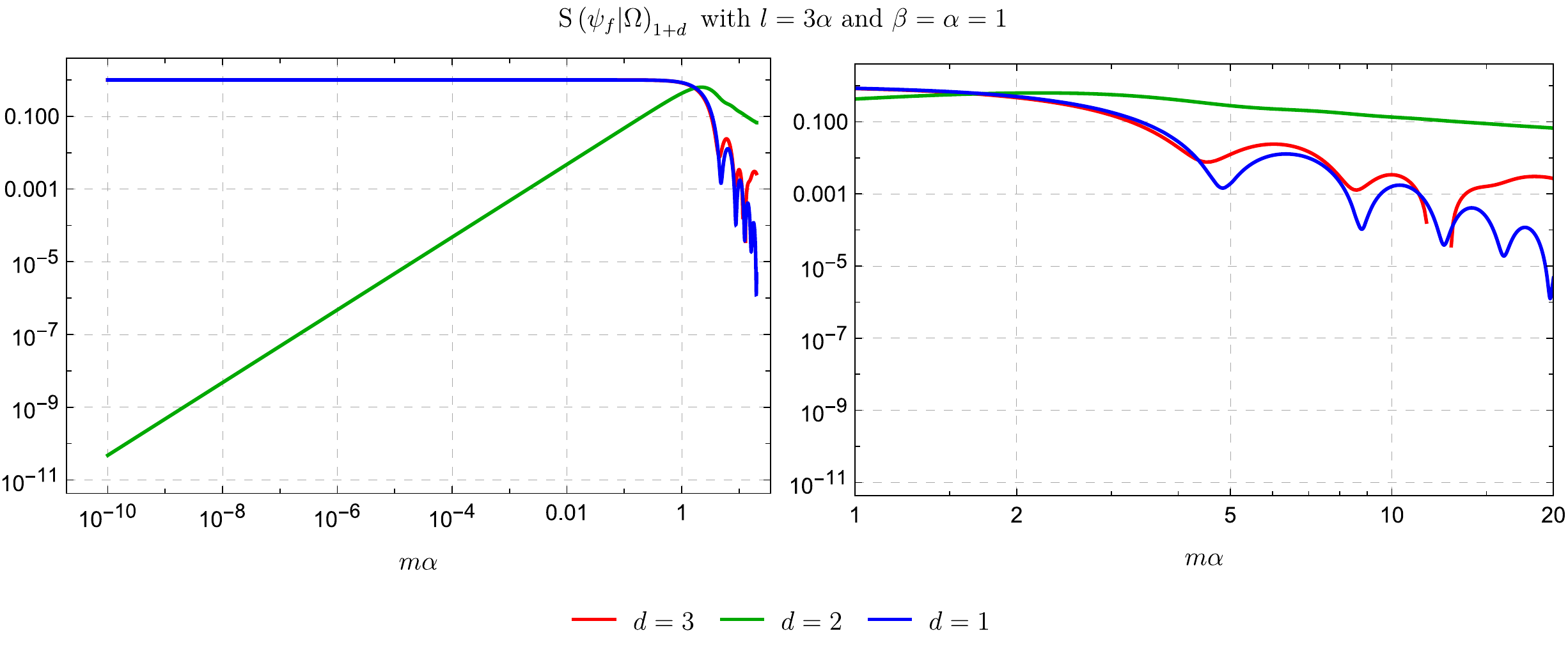}
		\caption{The log-log versions of the relative entropy plot from \Cref{fig:Results_1}. The left panel shows the relative entropy for $m\alpha \in [10^{-10},20]$ . The right panel shows it for $m\alpha \in [1,20]$. As can be seen in the left panel, the relative entropy for $d = 1,3$ varies very slowly until $m\alpha \approx 1$. On the other hand, the relative entropy for $d = 2$ presents a power-law increase. For larger values of $m\alpha$ the relative entropy  presents a power-law tail with visible oscillations for $m\alpha \gtrsim 5$ for $d = 1,3$. This decay is noticeably slower for $d = 2$ and has only one oscillation at $m\alpha \approx 7$ that is more visible in \Cref{fig:Results_1}. A discontinuity appears in the plot for $d = 3$ around $m\alpha \approx 10$ because the numerical value of the relative entropy becomes negative. This is not at odds with the positive-definiteness of the relative entropy because our result is still compatible with positive values, as can be seen in the error bands presented in \Cref{fig:Results_tail}.}
		\label{fig:Results_log}
	\end{figure*}

    \begin{figure*}
		\centering%
		\includegraphics[width=\textwidth]{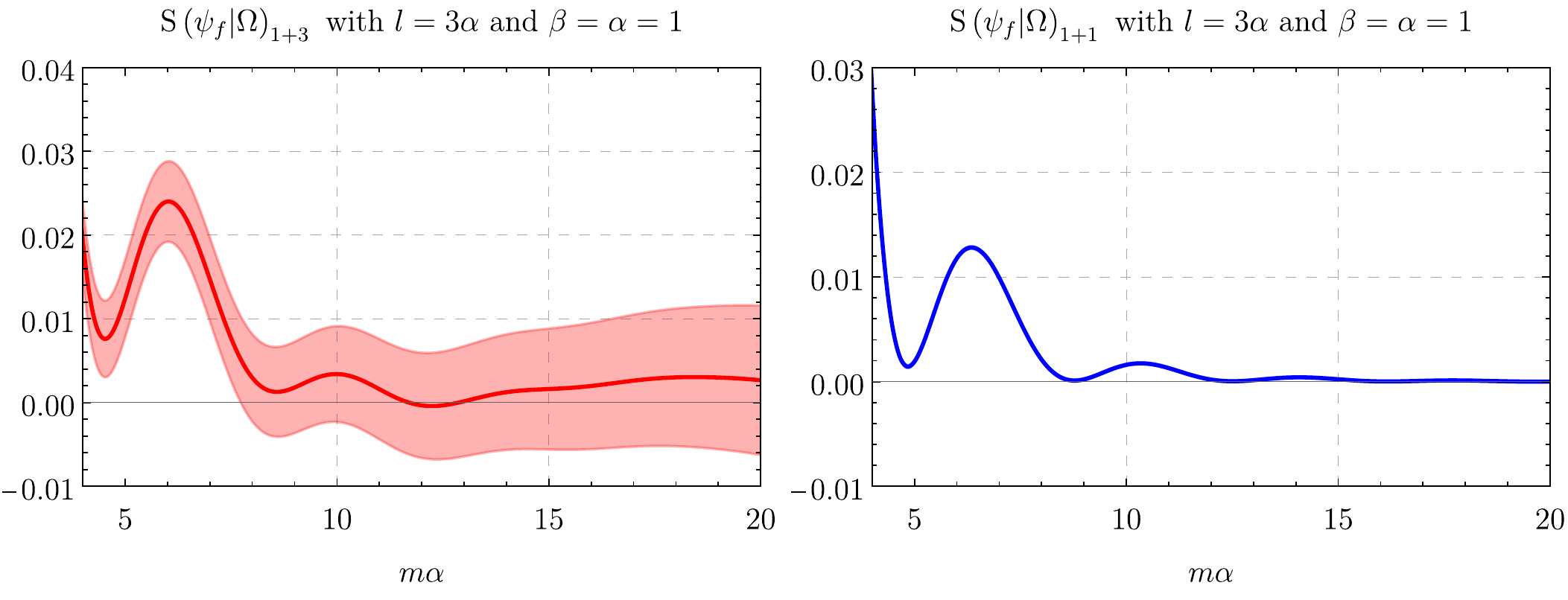}
		\caption{
        Linear-scale view of the $m\alpha \gtrsim 5$ region from \Cref{fig:Results_1}. Here the oscillations in the power-law decay of the relative entropy for $d = 3$ (left plot) and $d = 1$ (right plot) are more clearly visible. The red-shaded region for $d = 3$ represents the error bands in the relative entropy which were derived from the numerical integration error estimates. The error bands for $d = 1$ are not visible in this scale as the absolute error for this case is of order $\sim 10^{-4}$. For $m\alpha \gtrsim 8$, the relative entropy is comparable to the absolute error, and while the data remains consistent with a decaying, oscillatory tail, the detailed structure in this regime cannot be definitively resolved with our present numerical precision.
        }
		\label{fig:Results_tail}
	\end{figure*}

    This section contains our main findings about the dependence of the relative entropy on the mass and spacetime dimensionality. Using the numerical procedure described in Sec.~\labelcref{sec:Numerical_scheme} we computed the relative entropy with Eq.~\labelcref{ArakiSPJ} for $(1+d)$ spatial dimensions with $d\in\{1,2,3\}$.
    
    On general grounds, the correlations encoded in a given state of a scalar field play an important role in making them distinguishable from other states. This is particularly important for Gaussian states, which are completely characterized by the two-point correlations they encode. In this work we consider the vacuum state $\ket{\Omega}$ and a coherent state $\ket{\psi_f}$, both of which are examples of Gaussian states. Therefore, it is reasonable to expect that the difference in their two-point correlation structures within a given spacetime region governs their distinguishability.
    
    The mass $m$ controls how localized are these two-point correlations. In the large-mass regime $m\to\infty$, the typical wavelength associated with a given field mode becomes smaller. Consequently, the correlations of the field between a point $\vb*{x}$ and another point $\vb*{x'}$ are restricted to a neighborhood of $\vb*{x}$ that dwindles as $m\to\infty$. On the other hand, in the low-mass regime $m\to 0$, the opposite happens: the typical wavelength associated with a given field mode becomes arbitrarily large. Hence, the correlations of the field between a point $\vb*{x}$ and another point $\vb*{x'}$ extend to a neighborhood of $\vb*{x}$ that becomes larger as $m\to 0$. It is important to remark that the correlations of the field between a point $\vb*{x}$ and any other point $\vb*{x'}$ on its light-cone are agnostic to the mass $m$ due to the well-known singularities of the two-point functions on the light cone. Hence, in the large-mass limit, the relative importance of light-cone correlations grows, potentially leading to saturation effects in the distinguishability of $\ket{\Omega}$ and $\ket{\psi_f}$.
    
    This physical picture suggests that the relative entropy, and hence the distinguishability, between $\ket{\Omega}$ and $\ket{\psi_f}$ should monotonically decrease with increasing mass $m$. \Cref{fig:Results_1} confirms this expectation for $d=\{1,3\}$, where the relative entropy is a decreasing function of mass. However, the results for $d=2$, where the relative entropy is non-monotonic, initially increasing with mass, shows that such a reasoning is too simple. This highlights a non-trivial dependence on the dimensionality of the spacetime that cannot be captured by simple scaling arguments.

    The detailed behavior, as shown in \Cref{fig:Results_1}, further elucidates this dimensional dependence. In the low-mass regime ($m\alpha \ll 1$), the relative entropy for $d\in\{1,3\}$ peaks as $m \to 0$, while for $d=2$ it tends to zero. The behavior at exactly $m=0$ is singular: for $d=1$, the theory has a well-known infrared divergence, and for $d=2$, the relative entropy exhibits a discontinuity, jumping to a global maximum at $m=0$.
    
    As the mass increases through $m\alpha \approx 2$, a transition occurs. The concavity of the curve changes, and for $d=2$ the relative entropy reaches a local maximum near $m\alpha \approx 2.2$. For $m\alpha \gtrsim 5$, all cases enter a large-mass regime characterized by a power-law decay, upon which oscillatory behavior is superimposed.

    The high-mass regime is detailed in the log-log plot of \Cref{fig:Results_log} and the linear-scale zoom of \Cref{fig:Results_tail}. The power-law decay is noticeably slower for $d=2$ than for $d \in \{1,3\}$, for which the decay rates are similar. The oscillatory behavior also differs dimensionally: it is pronounced and multi-peaked for $d=1$ and $d=3$, but for $d=2$, it is rapidly damped, leaving only a faint oscillation visible around $m\alpha \approx 7$.

    For $d = 1$ we verified that the distances between its four visible successive peaks are
    \begin{equation}
        \begin{split}
            &\Delta m_{1\to 2} \approx 3.9/\alpha,\\
            &\Delta m_{2\to 3} \approx 3.8/\alpha\;\text{, and }\\
            &\Delta m_{3\to 4} \approx 3.6/\alpha,\\
        \end{split}
    \end{equation}
    where the notation $\Delta m_{i\to j}$ refers to the distance between the peaks $i$ and $j$, which we will henceforth abbreviate as peak-to-peak distance. For $d = 2$, we were not able to determine it because in this case there is only a faint oscillation in the relative entropy which is not enough to estimate the peak-to-peak distance. For $d = 3$ we are able to see only two successive peaks and the distance between them is $\Delta m_{1\to2} \approx 4.1/\alpha$. Given the similarity between the oscillatory character of the Pauli-Jordan distribution in this case and $d = 1$, it is reasonable to expect that $\Delta m_{i\to (i+1)}$ would also decrease for $d = 3$. To verify this conjecture, one would need to compute the relative entropy for $d = 3$ to a higher precision and possibly larger values of $m\alpha$ in order to obtain smaller error bands and perform a more rigorous analysis of the oscillations in the large-mass regime of the relative entropy.
    
    Heuristically, the oscillations in all cases can be understood in terms of the large-mass behavior of the Pauli-Jordan distribution (Eqs. ~\labelcref{eq::Paulis1,eq::Paulis2,eq::Paulis3}). For $d = 1$ and $d = 3$ it oscillates with a mass-dependent amplitude that is $\propto m^{-1/2}$ and $m^{1/2}$, respectively. Such a mass-dependent amplitude prevents the integration process from averaging the oscillations efficiently, which results in the oscillatory behavior we observed in the relative entropy. On the other hand, for $d = 2$ the Pauli-Jordan distribution oscillates with a mass-independent amplitude which allows the integration process to average them out, resulting in the lack of clear oscillations in this case.

    It is also possible to provide a rough estimate of the peak-to-peak distance for $d \in \{1,3\}$ using the typical length scale of the setup which is given by the edge $2\alpha = 2\beta$ of the region where the test function we used is supported. In the large-mass regime, the phase of the oscillations of the Pauli-Jordan distribution is, up to a constant factor
    \begin{equation}
        \phi = m\sqrt{(t-t')^2-(\vec{x}-\vec{x}')^2}.
    \end{equation}
    Considering the typical length scale of the problem, we can set \begin{equation}
        \sqrt{(t-t')^2-(\vec{x}-\vec{x}')^2} \approx 2\alpha.
    \end{equation} 
    This results in a typical peak-to-peak distance given by
    \begin{equation}
        \Delta m_{i\to i+1}2\alpha = 2\pi \Rightarrow \Delta m_{i\to i+1}\alpha= \pi \approx 3.14.
    \end{equation}
    As the reader might have noticed, this is smaller than the peak-to-peak distances observed in \Cref{fig:Results_tail}. This discrepancy arises because the estimate uses the large-mass asymptotics of the Pauli-Jordan distribution, which assumes $m \gg 1/(2\alpha)$. On the other hand, the plots considered masses up to $m\alpha = 20$ which is only a single order of magnitude larger than $1/2\alpha$. Therefore, the discrepancy is likely due to not being deep enough into the large-mass regime. The observation that $\Delta m_{i\to i+1}$ decreases as $m\alpha$ increases in $d = 1$ corroborates this hypothesis, which can be extended to the $d = 3$, given the similarity of the large-mass regime of the Pauli-Jordan distribution in these cases.

    In summary, our results demonstrate that the monotonic decay of the relative entropy with mass, previously observed in (1+1) dimensions~\cite{Nishioka:2018khk,Guimaraes:2025cqt}, is not a universal feature. The geometry of the underlying spacetime leaves an  imprint on the two-point functions of the field~\cite{Kempf_2021,Perche_2022}, which is inherited by the Pauli-Jordan distribution and, consequently, by the relative entropy. The non-monotonic behavior in $(1+2)$ dimensions and the dimension-dependent oscillatory tails reveal that spacetime dimensionality and the geometry of the smearing region play a fundamental role in the distinguishability of quantum states.

	%%%%%%%%%%%%%%%%%%%%%%%%%%%%%%%%%%
	\section{Conclusions} 
	\label{sec:Conclusions}
    
	In this work, we have explored the mass and spacetime dimensionality dependence of the relative entropy between a localized coherent state and the vacuum in a free scalar QFT, extending previous $(1+1)$-dimensional analyses to $(1+2)$ and $(1+3)$ dimensions. Our numerical results establish that the mass dependence exhibits striking dimensional sensitivity. In $(1+2)$ dimensions, the relative entropy is non-monotonic, initially increasing with mass before decaying, with only faint oscillatory structure. In $(1+3)$ and $(1+1)$ dimensions, clear oscillations emerge, decaying approximately as a power law. This demonstrates that dimensionality fundamentally modulates the mass dependence of the relative entropy through the Pauli-Jordan distribution, which is given by the geometry of the underlying spacetime.

    These findings advance the understanding of modular theory in relativistic QFT by explicitly quantifying how the informational distinguishability between the vacuum and localized excitations depends on the mass of the field and on the dimensionality of the underlying spacetime.

    Looking forward, several natural extensions emerge. First, it would be illuminating to study the combined $(m, \alpha)$ dependence in full parameter-space plots, in order to identify possible universal scaling regimes. Second, the generalization to interacting theories would probe how perturbative or nonperturbative dynamics deform the patterns established here. Third, connections to holography and black hole thermodynamics may be pursued, since the relative entropy plays a central role in modular energy inequalities and in the formulation of the quantum focusing conjecture \cite{Bousso:2015mna}. Finally, it would be interesting to extend the analysis beyond wedge regions to more general causal domains, thereby probing the interplay between geometry, mass, and quantum information in curved spacetime.

    In summary, our results demonstrate that the Araki-Uhlmann relative entropy serves as a sensitive measure of how spacetime dimensionality and field parameters govern quantum distinguishability. The observed dimensional dependence provides concrete data against which modular-theoretic and holographic proposals can be tested.
	
	%%%%%%%%%%%%%%%%%%%%%%%%%%
	\begin{acknowledgements}
		The authors would like to thank the Brazilian agencies CNPq, CAPES end FAPERJ for financial support.  S. P.~Sorella, I.~Roditi, and M. S.~Guimaraes are CNPq researchers under contracts 	302991/2024-7, 311876/2021-8, and 309793/2023-8, respectively. A. F. Vieira is supported by a postdoctoral grant from FAPERJ in the Pós-doutorado Nota 10 program,
grant No. 200.135/2025 and 200.136/2025.
	\end{acknowledgements}
		%%%%%%%%%%%%%%%%%%%%%%%%	
	%%%%%
    % Apêndices
    %%%%%
    \begin{widetext}
    \appendix
        \section{The light cone contribution to smeared Pauli-Jordan distribution in (1+3)D}\label{App: The light cone contribution to the PJ}
            On the $(1+3)$-dimensional Minkowski spacetime, consider a free massive scalar field. Its smeared Pauli-Jordan distribution is
                \begin{equation}\label{eq:smeared_pauli_jordan}
                    \Delta_{\rm PJ}(f,g) = \int \dd^{4}\vb*{x}\int\dd^{4}\vb*{x'}\, f(\vb*{x})\Delta_{\rm PJ}^{(1+3)}(\vb*{x}-\vb*{x'})g(\vb*{x'}),
                \end{equation}
                where $\vb*{x} = (t,\vec{x})$, $\vb*{x'} = (t',\vec{x}')$, $f,g\in\mathcal{C}_{0}^{\infty}(\mathbb{R}^4)$ are compactly supported smooth test functions, and the Pauli-Jordan distribution $\Delta_{\rm PJ}^{(1+3)}$ is given by equation~\labelcref{eq::Paulis3} which is composed by the sum of two terms, one supported when $\vb*{x}$ and $\vb*{x'}$ are timelike separated and the other one when $\vb*{x}$ and $\vb*{x'}$ are null separated. Consequently, we can write the smeared Pauli-Jordan distribution as
                \begin{equation}\label{eq:decomp-PJ}
                    \Delta_{\mathrm{PJ}}(f,g) = \Delta^{\mathrm{LC}}_{\mathrm{PJ}}(f,g) + \Delta^{\mathrm{BULK}}_{\mathrm{PJ}}(f,g),
                \end{equation}
                where
                \begin{equation}\label{eq: Null-PJ}
                    \Delta^{\rm LC}_{\rm PJ}(f,g) = -\frac{1}{4\pi}\int \dd^{3}\vec{x}\int\dd^{3}\vec{x}'\int\dd{t}\int\dd{t'}\, f(t,\vec{x})g(t',\vec{x}')\textmd{sign}(t-t')\delta((t-t')^2-R^2)
                \end{equation}
                is the contribution coming from the light cone and
                \begin{equation}\label{eq: BULK-PJ}
                    \Delta^{\rm BULK}_{\rm PJ}(f,g) = \frac{m}{8\pi}\int \dd^{3}\vec{x}\int\dd^{3}\vec{x}'\int\dd{t}\int\dd{t'}\, f(t,\vec{x})g(t',\vec{x}')\frac{J_1(m\sqrt{t^2-\vec{x}^{\,2}})}{\sqrt{t^2-\vec{x}^{\, 2}}}\textmd{sign}(t)\Theta(t^2-\vec{x}^{\,2}),
                \end{equation}
                is the contribution that arises from the bulk of the light cone. In both expressions, $R = |\vec{x}-\vec{x}'|$. 
                
                The light cone contribution admits further simplification by integrating over the Dirac delta distribution. By using the identity~\cite[Eq. 1.160]{Arfken_2013}
                \begin{equation}
                    \delta(h(x)) = \sum^{N}_{x_i} \frac{\delta (x - x_i)}{\abs{h'(x_i)}},
                \end{equation}
                where each $x_i$ is a root of the real-valued function $h(x)$, we can write
                \begin{equation}
                    \textmd{sign}(t-t')\delta((t-t')^2-(\vec{x}-\vec{x}')^{\,2}) = \frac{\delta(t'-t+R)}{2R} - \frac{\delta(t'-t-R)}{2R}.
                \end{equation}
                Plugging this result in equation~\labelcref{eq: Null-PJ} leads to
                \begin{equation}
                    \begin{split}
                        \Delta^{\rm LC}_{\rm PJ}(f,g) =& -\frac{1}{8\pi}\int \dd^{3}\vec{x}\int\dd^{3}\vec{x}'\frac{1}{R}\int\dd{t}\left(\int\dd{t'}\,f(t,\vec{x})g(t',\vec{x}')\delta(t'-t+R) - \right.
                        \\
                        &\left.\int\dd{t'}\, f(t,\vec{x})g(t',\vec{x}')\delta(t'-t-R)\right),
                    \end{split}
                \end{equation}
                which we integrate over $t'$ to obtain
                \begin{equation}
                    \begin{split}
                        \Delta^{\rm LC}_{\rm PJ}(f,g) =& \frac{1}{8\pi}\int \dd^{3}\vec{x}\int\dd^{3}\vec{x}'\frac{1}{R}\int\dd{t} f(t,\vec{x})\left(g(t+R,\vec{x}') - g(t-R,\vec{x}')\right).
                    \end{split}
                \end{equation}
    \end{widetext}
    
	\begingroup
	\allowdisplaybreaks
	%%%%%%%%%%%%%%%%%%%%%%%%%%%%%%%%%%
	
    \bibliographystyle{apsrev4-1}
%	\bibliographystyle{apsrev4-2}
	% Provoca muitos erros na compilação
	\bibliography{references}

@article{Guimaraes:2025cqt,
	author = "Guimaraes, Marcelo S. and Roditi, Itzhak and Sorella, Silvio P. and Vieira, Arthur F.",
	title = "{A numerical analysis of Araki-Uhlmann relative entropy in Quantum Field Theory}",
	eprint = "2502.09796",
	archivePrefix = "arXiv",
	primaryClass = "hep-th",
	doi = "10.1016/j.nuclphysb.2025.117011",
	journal = "Nucl. Phys. B",
	volume = "1018",
	pages = "117011",
	year = "2025"
}

@article{Tjoa:2021roz,
	author = "Tjoa, Erickson and Mart{\'\i}n-Mart{\'\i}nez, Eduardo",
	title = "{When entanglement harvesting is not really harvesting}",
	eprint = "2109.11561",
	archivePrefix = "arXiv",
	primaryClass = "quant-ph",
	doi = "10.1103/PhysRevD.104.125005",
	journal = "Phys. Rev. D",
	volume = "104",
	number = "12",
	pages = "125005",
	year = "2021"
}

@article{Caribe:2025knm,
    author = "Carib{\'e}, Jo{\~a}o G. A. and Guimaraes, Marcelo S. and Roditi, Itzhak and Sorella, Silvio P.",
    title = "{Investigating the Araki-Uhlmann relative entropy between two coherent states in relativistic Quantum Field Theory}",
    eprint = "2508.17165",
    archivePrefix = "arXiv",
    primaryClass = "hep-th",
    month = "8",
    year = "2025"
}

@book{Bogolyubov:1959bfo,
	author = "Bogolyubov, N. N. and Shirkov, D. V.",
	title = "{INTRODUCTION TO THE THEORY OF QUANTIZED FIELDS}",
	volume = "3",
	year = "1959"
}

@book{gradshteyn2014table,
	title={Table of integrals, series, and products},
	author={Gradshteyn, Izrail Solomonovich and Ryzhik, Iosif Moiseevich},
	year={2014},
	publisher={Academic press}
}

@article{Casini:2022rlv,
    author = "Casini, Horacio and Huerta, Marina",
    title = "{Lectures on entanglement in quantum field theory}",
    eprint = "2201.13310",
    archivePrefix = "arXiv",
    primaryClass = "hep-th",
    doi = "10.22323/1.403.0002",
    journal = "PoS",
    volume = "TASI2021",
    pages = "002",
    year = "2023"
}

@article{Hiai:1991mxv,
    author = "Hiai, Fumio and Petz, D\'enes",
    title = "{The proper formula for relative entropy and its asymptotics in quantum probability}",
    doi = "10.1007/BF02100287",
    journal = "Commun. Math. Phys.",
    volume = "143",
    number = "1",
    pages = "99--114",
    year = "1991"
}

@article{Marolf:2016dob,
    author = "Marolf, Donald and Wall, Aron C.",
    title = "{State-Dependent Divergences in the Entanglement Entropy}",
    eprint = "1607.01246",
    archivePrefix = "arXiv",
    primaryClass = "hep-th",
    doi = "10.1007/JHEP10(2016)109",
    journal = "JHEP",
    volume = "10",
    pages = "109",
    year = "2016"
}

@article{Holzhey:1994we,
    author = "Holzhey, Christoph and Larsen, Finn and Wilczek, Frank",
    title = "{Geometric and renormalized entropy in conformal field theory}",
    eprint = "hep-th/9403108",
    archivePrefix = "arXiv",
    reportNumber = "PUPT-1454, IASSNS-HEP-93-88",
    doi = "10.1016/0550-3213(94)90402-2",
    journal = "Nucl. Phys. B",
    volume = "424",
    pages = "443--467",
    year = "1994"
}

@article{Calabrese:2004eu,
    author = "Calabrese, Pasquale and Cardy, John L.",
    title = "{Entanglement entropy and quantum field theory}",
    eprint = "hep-th/0405152",
    archivePrefix = "arXiv",
    doi = "10.1088/1742-5468/2004/06/P06002",
    journal = "J. Stat. Mech.",
    volume = "0406",
    pages = "P06002",
    year = "2004"
}

@article{Calabrese:2009qy,
    author = "Calabrese, Pasquale and Cardy, John",
    title = "{Entanglement entropy and conformal field theory}",
    eprint = "0905.4013",
    archivePrefix = "arXiv",
    primaryClass = "cond-mat.stat-mech",
    doi = "10.1088/1751-8113/42/50/504005",
    journal = "J. Phys. A",
    volume = "42",
    pages = "504005",
    year = "2009"
}

@article{Berges:2017hne,
    author = {Berges, J\"urgen and Floerchinger, Stefan and Venugopalan, Raju},
    title = "{Dynamics of entanglement in expanding quantum fields}",
    eprint = "1712.09362",
    archivePrefix = "arXiv",
    primaryClass = "hep-th",
    doi = "10.1007/JHEP04(2018)145",
    journal = "JHEP",
    volume = "04",
    pages = "145",
    year = "2018"
}

@article{Schrofl:2023hnz,
    author = {Schr\"ofl, Markus and Floerchinger, Stefan},
    title = "{Relative Entropy and Mutual Information in Gaussian Statistical Field Theory}",
    eprint = "2307.15548",
    archivePrefix = "arXiv",
    primaryClass = "cond-mat.stat-mech",
    doi = "10.1007/s00023-024-01522-2",
    month = "7",
    year = "2023"
}

@article{Katsinis:2023hqn,
    author = "Katsinis, Dimitrios and Pastras, Georgios and Tetradis, Nikolaos",
    title = "{Entanglement of harmonic systems in squeezed states}",
    eprint = "2304.04241",
    archivePrefix = "arXiv",
    primaryClass = "hep-th",
    doi = "10.1007/JHEP10(2023)039",
    journal = "JHEP",
    volume = "10",
    pages = "039",
    year = "2023"
}

@article{Nishioka:2018khk,
    author = "Nishioka, Tatsuma",
    title = "{Entanglement entropy: holography and renormalization group}",
    eprint = "1801.10352",
    archivePrefix = "arXiv",
    primaryClass = "hep-th",
    reportNumber = "UT-18-02",
    doi = "10.1103/RevModPhys.90.035007",
    journal = "Rev. Mod. Phys.",
    volume = "90",
    number = "3",
    pages = "035007",
    year = "2018"
}

@article{Araki:1975zw,
    author = "Araki, H.",
    title = "{Inequalities in von Neumann Algebras}",
    year = "1975"
}

@article{Araki:1976zv,
    author = "Araki, H.",
    title = "{Relative Entropy of States of Von Neumann Algebras}",
    journal = "Publ. Res. Inst. Math. Sci. Kyoto",
    volume = "1976",
    pages = "809--833",
    year = "1976"
}

@article{Uhlmann:1976me,
    author = "Uhlmann, A.",
    title = "{Relative Entropy and the Wigner-Yanase-Dyson-Lieb Concavity in an Interpolation Theory}",
    doi = "10.1007/BF01609834",
    journal = "Commun. Math. Phys.",
    volume = "54",
    pages = "21",
    year = "1977"
}

@book{Takesaki:1970aki,
    author = "Takesaki, M.",
    title = "{Tomita's Theory of Modular Hilbert Algebras and its Applications}",
    doi = "10.1007/bfb0065832",
    publisher = "Springer-Verlag",
    series = "Lecture Notes in Mathematics",
    year = "1970"
}

@article{Witten:2018zxz,
    author = "Witten, Edward",
    title = "{APS Medal for Exceptional Achievement in Research: Invited article on entanglement properties of quantum field theory}",
    eprint = "1803.04993",
    archivePrefix = "arXiv",
    primaryClass = "hep-th",
    doi = "10.1103/RevModPhys.90.045003",
    journal = "Rev. Mod. Phys.",
    volume = "90",
    number = "4",
    pages = "045003",
    year = "2018"
}

@article{Casini:2019qst,
    author = "Casini, Horacio and Grillo, Sergio and Pontello, Diego",
    title = "{Relative entropy for coherent states from Araki formula}",
    eprint = "1903.00109",
    archivePrefix = "arXiv",
    primaryClass = "hep-th",
    doi = "10.1103/PhysRevD.99.125020",
    journal = "Phys. Rev. D",
    volume = "99",
    number = "12",
    pages = "125020",
    year = "2019"
}

@article{tomita1967canonical,
  title={On canonical forms of von Neumann algebras},
  author={Tomita, Minoru},
  journal={Fifth Functional Analysis Sympos (T{\^o}hoku Univ., Sendai, 1967), Math. Inst., Tohoku Univ., Sendai},
  pages={101--102},
  year={1967}
}

@article{Frob:2024ijk,
    author = {Fr\"ob, Markus B. and Sangaletti, Leonardo},
    title = "{Petz-R\'enyi relative entropy in QFT from modular theory}",
    eprint = "2411.09696",
    archivePrefix = "arXiv",
    primaryClass = "math-ph",
    month = "11",
    year = "2024"
}

@article{Hollands:2019czd,
    author = "Hollands, Stefan",
    title = "{Relative entropy for coherent states in chiral CFT}",
    eprint = "1903.07508",
    archivePrefix = "arXiv",
    primaryClass = "hep-th",
    doi = "10.1007/s11005-019-01238-z",
    journal = "Lett. Math. Phys.",
    volume = "110",
    number = "4",
    pages = "713--733",
    year = "2020"
}

@article{DAngelo:2021yat,
    author = "D'Angelo, Edoardo",
    title = "{Entropy for spherically symmetric, dynamical black holes from the relative entropy between coherent states of a scalar quantum field}",
    eprint = "2105.04303",
    archivePrefix = "arXiv",
    primaryClass = "gr-qc",
    doi = "10.1088/1361-6382/ac13b8",
    journal = "Class. Quant. Grav.",
    volume = "38",
    number = "17",
    pages = "175001",
    year = "2021"
}

@article{Ciolli:2021otw,
    author = "Ciolli, Fabio and Longo, Roberto and Ranallo, Alessio and Ruzzi, Giuseppe",
    title = "{Relative entropy and curved spacetimes}",
    eprint = "2107.06787",
    archivePrefix = "arXiv",
    primaryClass = "math-ph",
    doi = "10.1016/j.geomphys.2021.104416",
    journal = "J. Geom. Phys.",
    volume = "172",
    pages = "104416",
    year = "2022"
}

@article{Galanda:2023vjk,
    author = "Galanda, Stefano and Much, Albert and Verch, Rainer",
    title = "{Relative Entropy of Fermion Excitation States on the CAR Algebra}",
    eprint = "2305.02788",
    archivePrefix = "arXiv",
    primaryClass = "math-ph",
    doi = "10.1007/s11040-023-09464-7",
    journal = "Math. Phys. Anal. Geom.",
    volume = "26",
    number = "3",
    pages = "21",
    year = "2023"
}

@article{Garbarz:2022wxn,
    author = "Garbarz, Alan and Palau, Gabriel",
    title = "{Relative entropy of an interval for a massless boson at finite temperature}",
    eprint = "2209.00035",
    archivePrefix = "arXiv",
    primaryClass = "hep-th",
    doi = "10.1103/PhysRevD.107.125016",
    journal = "Phys. Rev. D",
    volume = "107",
    number = "12",
    pages = "125016",
    year = "2023"
}

@article{Floerchinger:2020ogh,
    author = "Floerchinger, Stefan and Haas, Tobias",
    title = "{Thermodynamics from relative entropy}",
    eprint = "2004.13533",
    archivePrefix = "arXiv",
    primaryClass = "cond-mat.stat-mech",
    doi = "10.1103/PhysRevE.102.052117",
    journal = "Phys. Rev. E",
    volume = "102",
    number = "5",
    pages = "052117",
    year = "2020"
}

@article{Jafferis:2014lza,
    author = "Jafferis, Daniel L. and Suh, S. Josephine",
    title = "{The Gravity Duals of Modular Hamiltonians}",
    eprint = "1412.8465",
    archivePrefix = "arXiv",
    primaryClass = "hep-th",
    reportNumber = "MIT-CTP-4611",
    doi = "10.1007/JHEP09(2016)068",
    journal = "JHEP",
    volume = "09",
    pages = "068",
    year = "2016"
}

@article{Jafferis:2015del,
    author = "Jafferis, Daniel L. and Lewkowycz, Aitor and Maldacena, Juan and Suh, S. Josephine",
    title = "{Relative entropy equals bulk relative entropy}",
    eprint = "1512.06431",
    archivePrefix = "arXiv",
    primaryClass = "hep-th",
    reportNumber = "NSF-KITP-15-162",
    doi = "10.1007/JHEP06(2016)004",
    journal = "JHEP",
    volume = "06",
    pages = "004",
    year = "2016"
}

@article{Verlinde:2019ade,
    author = "Verlinde, Erik and Zurek, Kathryn M.",
    title = "{Spacetime Fluctuations in AdS/CFT}",
    eprint = "1911.02018",
    archivePrefix = "arXiv",
    primaryClass = "hep-th",
    doi = "10.1007/JHEP04(2020)209",
    journal = "JHEP",
    volume = "04",
    pages = "209",
    year = "2020"
}

@article{Bousso:2020yxi,
    author = "Bousso, Raphael and Chandrasekaran, Venkatesa and Rath, Pratik and Shahbazi-Moghaddam, Arvin",
    title = "{Gravity dual of Connes cocycle flow}",
    eprint = "2007.00230",
    archivePrefix = "arXiv",
    primaryClass = "hep-th",
    doi = "10.1103/PhysRevD.102.066008",
    journal = "Phys. Rev. D",
    volume = "102",
    number = "6",
    pages = "066008",
    year = "2020"
}

@article{Dowling:2020nxc,
    author = "Dowling, Neil and Floerchinger, Stefan and Haas, Tobias",
    title = "{Second law of thermodynamics for relativistic fluids formulated with relative entropy}",
    eprint = "2008.02706",
    archivePrefix = "arXiv",
    primaryClass = "quant-ph",
    doi = "10.1103/PhysRevD.102.105002",
    journal = "Phys. Rev. D",
    volume = "102",
    number = "10",
    pages = "105002",
    year = "2020"
}

@article{Bousso:2015mna,
    author = "Bousso, Raphael and Fisher, Zachary and Leichenauer, Stefan and Wall, Aron C.",
    title = "{Quantum focusing conjecture}",
    eprint = "1506.02669",
    archivePrefix = "arXiv",
    primaryClass = "hep-th",
    doi = "10.1103/PhysRevD.93.064044",
    journal = "Phys. Rev. D",
    volume = "93",
    number = "6",
    pages = "064044",
    year = "2016"
}

@book{Arfken_2013,
    title = "Mathematical Methods for Physicists",
    year = {2013},
    booktitle = "Mathematical Methods for Physicists",
    author = "Arfken, George B. and Weber, Hans J. and Harris, Frank E.",
    publisher = {Elsevier},
    url = {https://linkinghub.elsevier.com/retrieve/pii/C20090306297},
    isbn = {9780123846549},
    doi = {10.1016/C2009-0-30629-7}
}

@misc{NIST:DLMF,
         key = "{\relax DLMF}",
       title = "{\it NIST Digital Library of Mathematical Functions}",
howpublished = "\url{https://dlmf.nist.gov/}, Release 1.2.1",
         url = "https://dlmf.nist.gov/",
        note = "F.~W.~J. Olver, A.~B. {Olde Daalhuis}, D.~W. Lozier, B.~I. Schneider,
                R.~F. Boisvert, C.~W. Clark, B.~R. Miller, B.~V. Saunders,
                H.~S. Cohl, and M.~A. McClain, eds.",
        year = "2024"
}

@book{Nestruev_2020,
  title={Smooth Manifolds and Observables},
  author={Nestruev, J.},
  isbn={9783030456498},
  lccn={2002026664},
  series={Graduate Texts in Mathematics},
  year={2020},
  publisher={Springer International Publishing}
}

@misc{Mathematica,
  author = {Wolfram Research{,} Inc.},
  title = {Mathematica, {V}ersion 13.1},
  note = {Champaign, IL, 2022}
}

@article{Perche_2022,
  title = {Geometry of spacetime from quantum measurements},
  author = {Perche, T. Rick and Mart\'{\i}n-Mart\'{\i}nez, Eduardo},
  journal = {Phys. Rev. D},
  volume = {105},
  issue = {6},
  pages = {066011},
  numpages = {21},
  year = {2022},
  month = {Mar},
  publisher = {American Physical Society},
  doi = {10.1103/PhysRevD.105.066011},
  url = {https://link.aps.org/doi/10.1103/PhysRevD.105.066011}
}

@article{Kempf_2021,
    AUTHOR={Kempf, Achim },
    TITLE={Replacing the Notion of Spacetime Distance by the Notion of Correlation},
    JOURNAL={Frontiers in Physics},
    VOLUME={Volume 9 - 2021},
    YEAR={2021},
    URL={https://www.frontiersin.org/journals/physics/articles/10.3389/fphy.2021.655857},
    DOI={10.3389/fphy.2021.655857},
    ISSN={2296-424X}
}
	
\end{document}